\documentclass[aps,prl,floatfix,twocolumn,byrevtex,superscriptaddress]{revtex4-1}
\usepackage{amsmath}
\usepackage{amssymb}
\usepackage{amstext}
\usepackage{amsopn}
\usepackage{amsfonts}
\usepackage{amsxtra}
\usepackage[english]{babel}
\usepackage{graphicx}
\usepackage{float}
\usepackage{bm}
\usepackage{multirow}
\usepackage{dcolumn}
\usepackage{color}
\usepackage{hyperref}

\newcommand{\icm}{\ensuremath{~\textrm{cm}^{-1}}}

\begin{document}

\title{Temperature-driven Topological Phase Transition and Intermediate Dirac Semimetal Phase in ZrTe$_5$}
\author{B. Xu}
\email[]{bing.xu@unifr.ch}
\affiliation{University of Fribourg, Department of Physics and Fribourg Center for Nanomaterials, Chemin du Mus\'{e}e 3, CH-1700 Fribourg, Switzerland}

\author{L. X. Zhao}
\affiliation{Beijing National Laboratory for Condensed Matter Physics, Institute of Physics, Chinese Academy of Sciences, P.O. Box 603, Beijing 100190, China}

\author{P. Marsik}
\author{E. Sheveleva}
\author{F. Lyzwa}
\affiliation{University of Fribourg, Department of Physics and Fribourg Center for Nanomaterials, Chemin du Mus\'{e}e 3, CH-1700 Fribourg, Switzerland}

\author{Y. M. Dai}
\affiliation{Center for Superconducting Physics and Materials, National Laboratory of Solid State Microstructures and Department of Physics, Nanjing University, Nanjing 210093, China}

\author{G. F. Chen}
\affiliation{Beijing National Laboratory for Condensed Matter Physics, Institute of Physics, Chinese Academy of Sciences, P.O. Box 603, Beijing 100190, China}

\author{X. G. Qiu}
\affiliation{Beijing National Laboratory for Condensed Matter Physics, Institute of Physics, Chinese Academy of Sciences, P.O. Box 603, Beijing 100190, China}

\author{C. Bernhard}
\affiliation{University of Fribourg, Department of Physics and Fribourg Center for Nanomaterials, Chemin du Mus\'{e}e 3, CH-1700 Fribourg, Switzerland}

\date{\today}
%
%

\begin{abstract}
We present an infrared spectroscopy study of ZrTe$_5$, which confirms a recent theoretical proposal that this material exhibits a temperature-driven topological quantum phase transition from a weak to a strong topological insulating state with an intermediate Dirac semimetal state around $T_p \simeq$ 138~K. Our study details the temperature evolution of the energy gap in the bulk electronic structure. We found that the energy gap closes around $T_p$ where the optical response exhibits characteristic signatures of a Dirac semimetal state, i.e. a linear frequency-dependent optical conductivity extrapolating to the origin (after subtracting a weak Drude response). This finding allows us to reconcile previous diverging reports about the topological nature of ZrTe$_5$ in terms of a variation of $T_p$ that depends on the crystal growth condition.
\end{abstract}



\maketitle

%
%
The discovery of topological insulators (TIs), materials that are insulators in the bulk, while having metallic surface states connecting the bulk valence and conduction bands, has stimulated a very intensive search for novel topological phases of matter~\cite{Hasan2010RMP}. In the past few years, many topological quantum materials have been predicted and experimentally confirmed, including three-dimensional (3D) Dirac semimetals and Weyl semimetals~\cite{Liu2014SC,Liu2014NM,Weng2015PRX,Huang2015PRX,Xu2015SC,Lv2015PRX}. Recently, the layered transition-metal pentatelluride ZrTe$_5$, a material that has been studied for decades due to its large thermoelectric power and a mysterious resistivity anomaly~\cite{Okada1980,Jones1982,Tritt1999}, has ignited renewed interest as a candidate of a novel topological material. A monolayer of ZrTe$_5$ was predicted to be a good quantum spin Hall insulator with a large bulk band gap~\cite{Weng2014PRX}. Furthermore, it was argued that bulk ZrTe$_5$ can be either in a weak or strong 3D TI state of which only the latter has topologically protected surface states~\cite{Weng2014PRX}. Notably, it was predicted that a transition between these states can be induced by a small change of the lattice parameters, for example as the temperature is varied. Such a topological phase transition (TPT) involves an inversion of the valence and conduction bands and thus an intermediate state at which the band gap closes and a 3D Dirac semimetal phase may arise~\cite{Manzoni2016PRL,Fan2017}.

The experimental search for a TPT in ZrTe$_5$ seems to be complicated by its strong sensitivity on the details of the lattice parameter and probably also the purity of the crystals. As a result, the experimental results seem to be fairly contradictory. Several angle-resolved photoemission spectroscopy (ARPES) studies~\cite{Li2016NP,Shen2017} and the chiral magnetic effect in ZrTe$_5$~\cite{Li2016NP} suggest a 3D Dirac semimetal band structure without a finite band gap. Infrared spectroscopy (IR)~\cite{Chen2015PRB}, magneto-optical~\cite{Chen2015PRL} and transport~\cite{Zheng2016PRB} measurements were also interpreted in terms of a 3D Dirac semimetal. On the other hand, two recent scanning tunneling microscopy (STM) studies found a large bulk band gap of about 80 to 100~meV and concluded that ZrTe$_5$  is a weak 3D TI~\cite{Li2016PRL,Wu2016PRX}. A recent high-resolution laser based APRES study also favours a weak TI state~\cite{Moreschini2016PRB}. To the contrary, other ARPES and STM studies revealed a metallic character of the sample surface and argued that ZrTe$_5$ is a strong TI~\cite{Manzoni2016PRL,Manzoni2017}.

An important clue for resolving this puzzle emerges from a recent magneto-transport study which indicates that the TPT coincides with a resistivity peak~\cite{Nair2018PRB} that is known to be sample dependent with a variation of the reported peak temperatures, $T_p$, from about 10 -- 160~K~\cite{Nair2018PRB,Chi2017,Manzoni2016PRL}. This suggests that the complete TPT transition, in particular, the reopening of the inverted band gap in the strong TI state at low temperature, can only be observed for samples with sufficiently high $T_p$ values. In particular, since the sample of a previous IR study had a $T_p$ value of only 50~K, this calls for a reinvestigation of the IR response with a sample that has a substantially higher $T_p$. In this letter we present such a temperature dependent IR study of a high-quality single crystal of ZrTe$_5$ with a resistivity peak at $T_p$ $\simeq$ 138~K. We observe characteristic signatures of a TPT with an intermediate Dirac semimetal state around $T_p$ that are consistent with the theoretical predictions and the present experimental results of ZrTe$_5$.

%
%
\begin{figure*}[tb]
\includegraphics[width=0.98\textwidth]{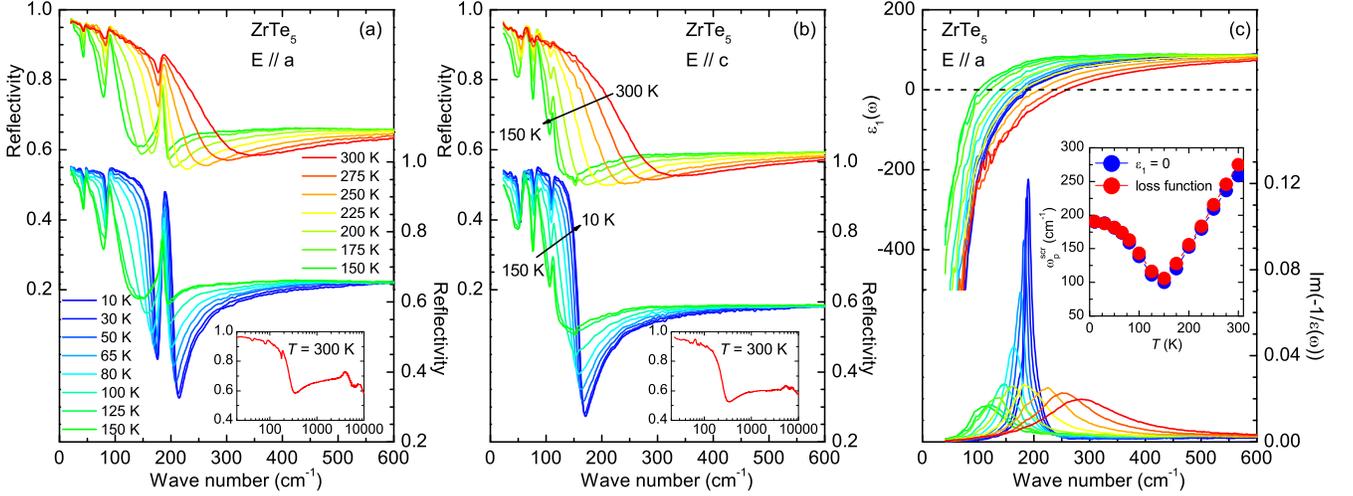}
\caption{ (color online) Far-infrared reflectivity of ZrTe$_5$ along the $a$-axis (panel a) and $c$-axis (panel b) at high (top) and low (bottom) temperatures. The insets show the room temperature spectrum up to 10\,000\icm. (c) Temperature dependence of the real part of the dielectric function $\varepsilon_1(\omega)$ (top) and the loss function Im($-1/\varepsilon(\omega)$) (bottom) along the $a$-axis, after the contribution of the phonon modes has been subtracted. Inset shows the screened plasma frequency of the free carries obtained from the zero crossings of $\varepsilon_1(\omega)$ (blue circles) and the peak of the loss function (red circles).}
\label{Fig1}
\end{figure*}
\begin{figure*}[tb]
\includegraphics[width=0.98\textwidth]{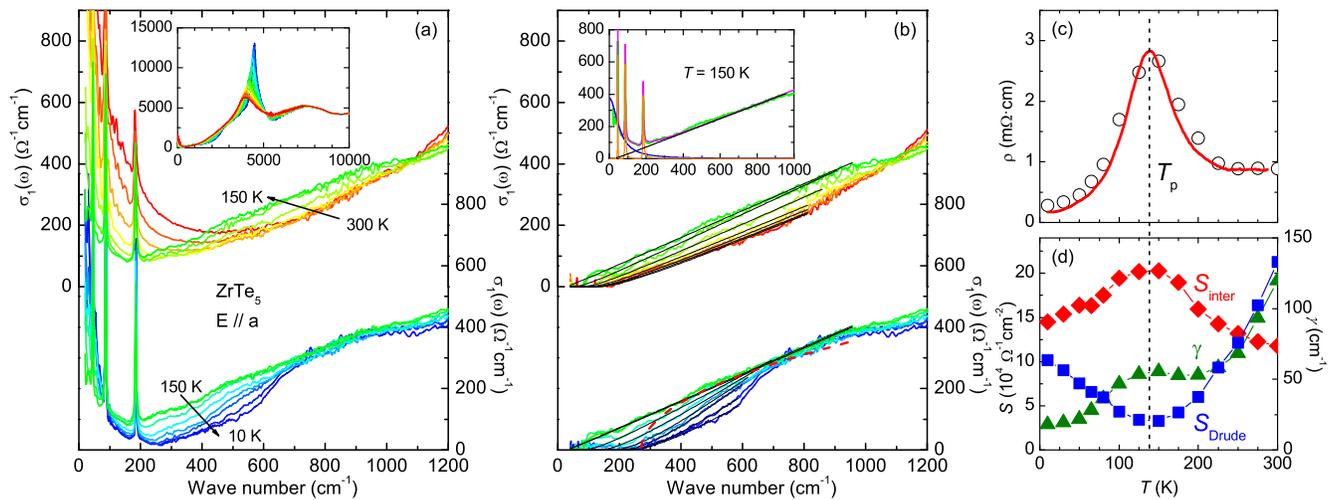}
\caption{ (color online) (a) Optical conductivity $\sigma_1(\omega)$ of ZrTe$_5$ along the $a$-axis at high (top) and low (bottom) temperatures. Inset displays $\sigma_1(\omega)$ up to 10\,000\icm. (b) Optical conductivity spectra along the $a$-axis, after the Drude response and the phonon modes have been subtracted. Inset: Fit of $\sigma_1(\omega)$ at 150~K; thin solid lines represent the Drude (blue), phonon modes (orange), and interband (black) terms. (c) Comparison of the dc resistivity, $\rho_{a}$ (solid line), with the zero-frequency values of the Drude fits to the conductivity (open circles). (d) Temperature dependences of the Drude weight (blue squares), the interband spectral weight (red diamonds) and the electronic scattering rate (green triangles).
}
\label{Fig2}
\end{figure*}

High-quality single crystals of ZrTe$_5$ have been synthesized with a chemical vapor transport method. They exhibit a pronounced resistivity peak at $T_p$ $\simeq$ 138~K [see e.g. Fig.~\ref{Fig2}(c)] in good agreement with most previous studies~\cite{Okada1980,Jones1982,Manzoni2016PRL,Zhang2017PRL,Zhang2017NC,Xiong2017PRB,Zhou2016PNAS,Yuan2016}. The details of the IR reflectivity measurements and the Kramers-Kronig analysis are described in the Supplementary Material~\footnote{See Supplemental Material at \url{http://link.aps.org/supplemental/xxx} for the reflectivity measurements and the Kramers-Kronig analysis, which includes Refs.~\cite{Homes1993,Dressel2002}}.

%
%
The temperature dependences of the far-infrared reflectivity $R(\omega)$ of ZrTe$_5$ along the $a$-axis and $c$-axis directions is shown in Fig.~\ref{Fig1}(a) and Fig.~\ref{Fig1}(b), respectively. The inset in each panel shows the room temperature spectrum up to 10\,000\icm. In the low-frequency region, $R_{a}(\omega)$ shows a typical metallic response with a rather sharp plasma edge below which it rapidly approaches unity. In addition, we identify three sharp phonon peaks around 45, 85 and 185\icm. The plasma edge in $R_{a}(\omega)$ exhibits a rather strong and highly unusual $T$-dependence, since it decreases at first from about 300\icm\ at 300~K to about 100\icm\ at 150~K and then increases again at lower temperature to about 200\icm\ at 10~K. A nearly identical behaviour of the plasma edge is seen in the $c$-axis response in Fig.~\ref{Fig1}(b), for which the major difference concerns the position of the phonon modes at about 60, 80 and 110\icm. In the following we focus on the characteristic features along the $a$-axis.

The top panel of Fig.~\ref{Fig1}(c) shows the real part of the $a$-axis dielectric function $\varepsilon_1(\omega)$, where the contribution of the sharp phonon modes has been subtracted to emphasize the zero crossing of $\varepsilon_1(\omega)$. At low frequencies, $\varepsilon_1(\omega)$ is negative, indicating a defining property of a metal, which can be described by the Drude model $\varepsilon(\omega) = \varepsilon_{\infty} - \omega_p^2/(\omega^2 + i\omega\gamma)$, where $\varepsilon_{\infty}$ is the high-energy dielectric constant, $\omega_p$ the Drude plasma frequency, and $\gamma$ the scattering rate. The high value of $\varepsilon_{\infty}$ of about 80--90 contributes to the rather low value of the zero crossing of $\varepsilon_1(\omega)$ for which the $T$-dependence is detailed in the inset (blue circles). The latter corresponds to the screened Drude plasma frequency $\omega_p^{scr}$ of the free carriers, which is related to $\omega_p$ through $\omega_p^{scr} = \omega_p/\sqrt{\varepsilon_{\infty}}$. The temperature dependence of $\omega_p^{scr}$ agrees indeed well with the minimum of the plasma edge of $R_{a}(\omega)$ in Fig.~\ref{Fig1}(a). Moreover, the $T$-dependence of the longitudinal response of this material, namely the loss function Im($-1/\varepsilon(\omega)$), is plotted in the bottom panel of Fig.~\ref{Fig1}(c). As shown by the red circles in the inset of Fig.~\ref{Fig1}(c), the peak in Im($-1/\varepsilon(\omega)$) is also a good estimate of $\omega_p^{scr}$. All three parameters consistently show that $\omega_p^{scr}$ first decreases from about 300\icm\ at 300~K toward a sharp minimum of 100\icm\ at 140~K, i.e. in the vicinity of $T_p$, and then increases again at lower temperature where it saturate around 200\icm.

Figure~\ref{Fig2}(a) displays the $T$-dependence of $\sigma_1(\omega)$ in the far-infrared range. At low frequencies, the free-carrier contribution to $\sigma_1(\omega)$ is seen as a Drude-like peak. Upon cooling, the Drude peak becomes narrower and its spectral weight initially decreases from 300 to 150~K and then increases again from 150 to 10~K. More quantitatively, as shown by the inset of Fig.~\ref{Fig2}(b), we can fit the low-energy $\sigma_1(\omega)$ with a Drude model. Fig.~\ref{Fig2}(c) shows the dc resistivity $\rho \equiv 1/\sigma_{1}(\omega = 0)$, derived from the fitted zero-frequency value (open circles), which agrees well with the transport data (solid curve), indicating that our modeling is reliable. The corresponding Drude spectral weight, $S_{Drude} = \frac{\pi^2}{Z_0}\omega_p^2$, where $Z_0$ is the vacuum impedance, and the scattering rate, $\gamma$, are displayed in Fig.~\ref{Fig2}(d). The $T$-dependence of $S_{Drude}$ agrees well with the one of $\omega_p^{scr}$ as discussed above and shown in Fig.~\ref{Fig1}(c). Moreover, the minimum of $S_{Drude}$ coincides with $T_p$ $\simeq$ 138~K.

The inset of Fig.~\ref{Fig2}(a) shows the $\sigma_1(\omega)$ spectra up to 10\,000\icm. It reveals a sharp peak around 4\,500\icm\ that was also observed in Ref.~\cite{Chen2015PRB} and interpreted in terms of a Van Hove singularity in the joint density of state.

Next, we investigate the signature of the low-energy interband transitions. To single out this contribution we show in Fig.~\ref{Fig2}(b) the $\sigma_1(\omega)$ spectra, after the Drude response and the sharp phonon modes have been subtracted. An interesting feature is the almost perfect linearity of $\sigma_1(\omega)$ in the low frequency regime (up to about 900\icm\ at 150~K), which is a strong indication for the presence of 3D linear bands near the Fermi level~\cite{Chen2015PRB}. The onset of the interband transitions can be estimated by a linear extrapolation of the conductivity to $\sigma_1(\omega) = 0$. This intercept is located around 275\icm\ (34~meV) at 10~K, and decreases to around 32\icm\ (4~meV) at 150~K, before it increases again to about 184\icm\ (23~meV) at 300~K.

\begin{figure}[tb]
\includegraphics[width=0.95\columnwidth]{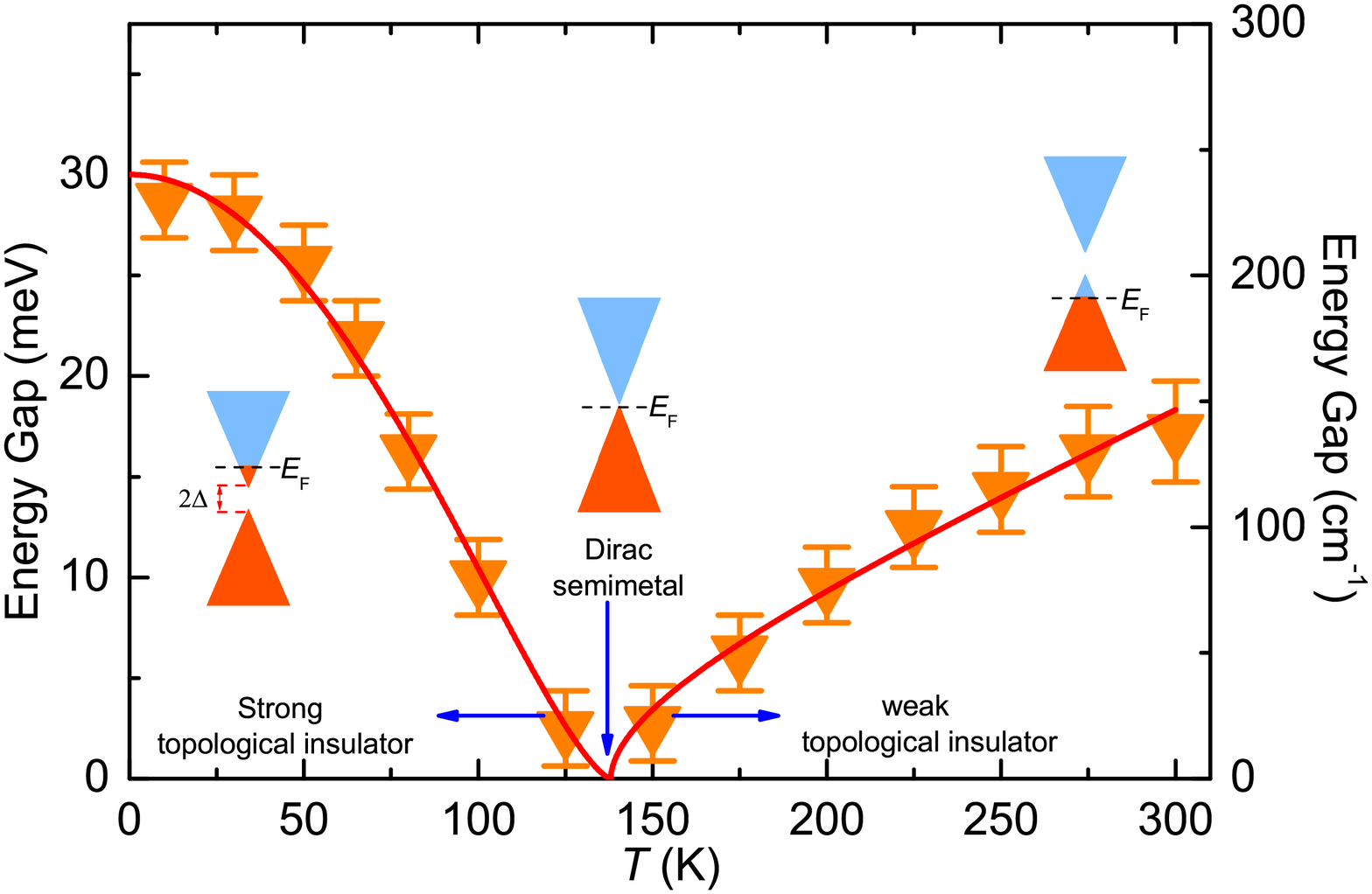}
\caption{ (color online) Experimentally obtained value of the band gap for ZrTe$_5$ at different temperatures. The solid lines are guides to the eye. The diagrams schematically show the temperature evolution of the bulk band at the $\Gamma$ point.
}
\label{Fig3}
\end{figure}

For a more detailed analysis of the linearity of $\sigma_1(\omega)$, we recall the theoretically expected behaviour of 3D Dirac fermions. In the simplest case where the Dirac point is located right at the Fermi level, the interband conductivity is given by $\sigma_1(\omega) = \frac{NG_0}{24}\frac{\omega}{\nu_F}$, where $N$ is the number of Weyl points, $G_0 = 2e^2/h$ the quantum conductance and $\nu_F$ the Fermi velocity. If the Fermi level is not at the Dirac point ($E_F \neq 0$), $\sigma_1(\omega) = \frac{NG_0}{24}\frac{\omega}{\nu_F}\Theta(\omega - 2|E_F|)$, where the linear part of $\sigma_1(\omega)$ can still be extrapolated to the origin, but the interband transitions will be terminated below $\omega = 2|E_F|$, due to the blocking of the electronic states. The latter are seen as a Drude peak that accounts for the intraband transitions of the Dirac cone~\cite{Hosur2012PRL,Tabert2016PRB,Ashby2014,Timusk2013PRB,Xu2016PRB,Xu2017NC,Tabert2016}. If the Dirac cone has a massless gap, 2$\Delta$, and $E(k) = \pm(\hbar\nu_Fk + \Delta)$, the interband conductivity becomes~\cite{Tabert2016,Neubauer2016,Benfatto2008}
%
\begin{equation}
\sigma_{1}(\omega) = \frac{N G_{0}}{24v_{F}}\frac{(\omega - 2\Delta)^2}{\omega}\Theta(\omega - \mathrm{max}[2|E_F|, 2\Delta]),
\label{SigmaDirac}
\end{equation}
where the extrapolation of the interband conductivity has a finite intercept that is associated with the energy gap. For the scenario of a massive gap~\cite{Tabert2016,Chen2017PNAS}, where $E(k) = \pm\sqrt{(\hbar\nu_Fk)^2 + \Delta^2}$, the interband conductivity follows $\sigma_1(\omega) \propto \sqrt{\omega - 2\Delta}$ near $\omega \sim 2\Delta$ [indicated by the red dashed curve in Fig.~\ref{Fig2}(b)], and the linear part of $\sigma_1(\omega)$ at $\omega \gg 2\Delta$ can still be extrapolated to the origin. Our optical data therefore are rather consistent with the massless gap model. As shown by the thin solid lines through the data in Fig.~\ref{Fig2}(b), we can reproduce the low-energy interband conductivity by using Eq.~\ref{SigmaDirac}, and the temperature dependence of the obtained gap parameter is displayed in Fig.~\ref{Fig3}.

Accordingly, the following sequence naturally explains the $T$-dependence of the Dirac energy gap in ZrTe$_5$. Starting from 10~K, as the temperature is increased, the band gap first decreases gradually and vanishes when the valence and conduction bands touch at the critical temperature $T_p$ $\simeq$ 138~K. At this point a Dirac semimetal state is formed whose characteristic signatures are the above mentioned linearity of $\sigma_1(\omega)$ and its extrapolation to $\sigma_1(\omega) = 0$ at zero frequency. As the temperature is further increased, the band gap opens again, and ultimately reaches about 20~meV at 300~K. This temperature evolution of the band topology is schematically illustrated in Fig.~\ref{Fig3}. Theoretically, the Dirac semimetal state of ZrTe$_5$ is realized at the boundary between the strong and weak TI states~\cite{Manzoni2016PRL,Fan2017}. In this context, our data can confirm a temperature driven transition between the strong and weak TI states across a critical gapless state at $T_p$. Accordingly, the topologically protected strong TI state is at the low temperature side, while the weak TI state is at the high temperature side. The temperature mainly affects the band structure and the energy gap at the $\Gamma$ point of the Brillouin zone. Furthermore, our observations suggest that away from this phase boundary, the range of the linear band dispersion is shrinking. This is evident in Fig.~\ref{Fig2}(b) from the interband-conductivity spectra for which the range with a linear frequency dependence decreases as the temperature is increased or decreased away from $T_p$.

This closing and reopening of the band gap also causes a characteristic spectral weight transfer between the intraband and interband transitions. As the gap closes, the onset frequency of the interband transitions moves to lower frequency, giving rise to the enhanced interband spectral weight, and vice versa for the reopening of the gap. Fig.~\ref{Fig2}(d) depicts the temperature dependence of the interband spectral weight, $S_{inter} = \int_{0}^{\omega_c}\sigma_{1}(\omega)d\omega$, where $S_{inter}$ is calculated by integrating the interband $\sigma_{1}(\omega)$ shown in Fig.~\ref{Fig2}(b) from 0 to 1\,000\icm. It shows that the temperature dependences of $S_{inter}$ and $S_{Drude}$ follow opposite trends with a maximum, respectively a minimum around $T_p$ $\simeq$ 138~K. The characteristic spectral weight transfer from the interband to the intraband transitions as one moves away from $T_p$, either to lower or higher temperature, is also consistent with the scenario that $E_F$ moves into the conduction/valence band, because if $E_F$ locates at the gap region, i.e. $\Delta > E_F$, then $S_{Drude}$ at the gap-closing temperature $T_p$ would reach a maximum instead of the observed minimum. Such a $T$-induced Lifshitz transition from a low-$T$ state with an electron-pocket to a high-$T$ state with a hole-like Fermi surface was indeed observed in a recent ARPES study~\cite{Zhang2017NC}, as well as the previous bulk Hall coefficient and thermopower measurements~\cite{Jones1982,Izumi1982,McIlroy2004}. The charge neutrality point ($E_F = 0$) should be around $T_p$, where the small residual Drude weight can be accounted for in terms of thermal excitations and a possible broadening of the transition due to some sample inhomogeneity. Considering the effect of thermal excitations, one can also understand that the Drude peak acquires more spectral weight on the high temperature side of the transition. In addition, an interesting question is how temperature can cause a significant charge carrier change and the related Fermi level shift without obvious charge carrier doping. Several possibilities have been discussed in the recent ARPES study of ZrTe$_5$~\cite{Zhang2017NC} that are compatible with our optical data.

The Fermi level shift caused by temperature also complicates the analysis of the onset of the interband transitions to identify the Dirac energy gap. If $E_F$ is away from the charge neutrality point, i.e. $E_F > \Delta$, the energy of the interband transitions onset characterizes $2|E_F|$ instead of $2\Delta$. The energy gap values in Fig.~\ref{Fig3} thus can be viewed as an upper limit of the bulk band gap which ranges from about 30~meV for the strong TI at 10~K to 20~meV for the weak TI at 300~K. This assignment agrees with a recent theoretical work~\cite{Fan2017} which predicts a gap of $-$33~meV and 28~meV in the strong and weak TI phase, respectively. It is also compatible with recent high-resolution ARPES studies which find a strongly temperature dependent energy gap with a maximum of about 30~meV~\cite{Zhang2017NC,Xiong2017PRB}.

Finally, we comment on the importance of our observations. Generally, in a non-interacting system a band inversion is the hallmark of a TPT, where the bulk band gap closes at the phase transition and reopens with inverted characters of the bottom conduction and top valence bands~\cite{Zhang2009NP}. Accordingly, our observation of a closing and reopening of the bulk band gap with an intermediate 3D Dirac semimetal state at $T_p$ supports the scenario of a temperature-induced TPT in ZrTe$_5$, that is driven by the tuning of the inter-layer distance~\cite{Manzoni2016PRL} or the volume expansion~\cite{Fan2017}. Our observations, as summarized in the phase diagram in Fig.~\ref{Fig3}, suggest an intrinsic connection between the TPT and the anomaly of the dc transport at $T_p$ that is very sensitive to the change of the crystal volume~\cite{Weng2014PRX,Manzoni2016PRL,Fan2017} and can be modified by temperature, pressure~\cite{Zhou2016PNAS,Zhang2017PRL}, and even the growth technique and sample quality~\cite{Manzoni2016PRL,Nair2018PRB,Chi2017,Zhang2017PRL,Zhang2017NC,Xiong2017PRB,Yuan2016}. Fortunately, our ZrTe$_5$ crystal has a high value of $T_p$ $\simeq$ 138~K that allows all three states, i.e. the strong TI at $T \ll T_p$, the 3D Dirac semimetal at $T \simeq T_p$ and the weak TI at $T \gg T_p$, to be readily studied. Considering the variation of $T_p$, one can also reconcile the contradicting experimental reports about the topological nature of ZrTe$_5$, which vary between a Dirac semimetal~\cite{Li2016NP,Chen2015PRB,Chen2015PRL,Zheng2016PRB}, a strong or weak topological insulator~\cite{Manzoni2016PRL,Fan2017,Li2016PRL,Wu2016PRX,Chen2017PNAS}, and a gapped semiconductor~\cite{Moreschini2016PRB}. In the near future, this unusual flexibility of the topological properties of ZrTe$_5$ may also find applications in various electronic devices.

%
%
To summarize, we have studied the frequency dependent optical conductivity of ZrTe$_5$ at temperatures between 10 and 300~K. Our optical data provide clear evidence for the closing and reopening of the bulk band gap as a function of temperature with an intermediate gapless state that coincides with a peak in the resistivity at $T_p$ $\simeq$ 138~K. In the latter state, the optical conductivity (after substraction of a weak Drude peak) shows the characteristic signatures of a 3D Dirac semimetal in terms of a linear frequency dependence between 0 and about 100~meV and an intercept at the origin of the frequency axis. Our optical data are consistent with the theoretically predicted topological phase transition from a strong TI state to a weak TI state with an intermediate Dirac semimetal state. They also agree with most of the published experimental work if one takes into account that $T_p$ can be sample dependent and vary between 10 and 160~K. They may be a useful guide for further studies of the Dirac semimetal state at $T_p$ and the topological surface states which should be observable only in the strong TI state at $T \ll T_p$.

%
%
\begin{acknowledgments}
We acknowledge very illuminating discussions with Binghai Yan. Work at the University of Fribourg was supported by the Schweizer Nationalfonds (SNF) by Grant No. 200020-172611.
\end{acknowledgments}

%

\end{document}